%% file: dphi.tex
\documentclass[aps,prl,twocolumn,showpacs,superscriptaddress,groupedaddress]{revtex4}  
\usepackage{graphicx}  
\usepackage{dcolumn}   
\usepackage{bm}        
\usepackage{amssymb}   
\usepackage{slashed}
\usepackage{dcolumn}

\newcommand{\Rdphi}{R_{\Delta \phi}}

\newcommand{\Dphimax}{\Delta \phi_{\rm max}}
\newcommand{\Dphi}{\Delta \phi_{\rm dijet}}

\newcommand{\ptmin}{p_{T \rm min}}

\newcommand{\ystar}{y^*}
\newcommand{\ystarmax}{y^*_{\rm max}}
\newcommand{\yboost}{y_{\rm boost}}

\newcommand{\mur}{\mu_R}
\newcommand{\muf}{\mu_F}

\newcommand{\as}{\alpha_s}

\newcommand{\asmz}{\alpha_s(M_Z)}

\newcommand{\pythia}{{\sc pythia}}
\newcommand{\herwig}{{\sc herwig}}
\newcommand{\sherpa}{{\sc sherpa}}

\newcommand{\Rcone}{R_{\rm cone}}

\newcommand{\ord}{{\cal O}}
\newcommand{\ppbar}{p{\bar{p}}}
\newcommand{\epem}{e^+e^-}
\newcommand{\Ndof}{N_{\rm dof}}

\begin{document}

\hspace{5.2in} \mbox{FERMILAB-PUB-12-650-E}

\title{\boldmath
Measurement of the combined rapidity and $p_T$ dependence 
of dijet azimuthal decorrelations
in $p\bar{p}$ collisions at $\sqrt{s}=1.96\,$TeV}

\input author_list.tex

\date{December 8, 2012}

\begin{abstract}
We present the first combined measurement of the rapidity and transverse 
momentum dependence of dijet azimuthal decorrelations, using the 
recently proposed quantity $\Rdphi$.
The variable $\Rdphi$ measures the fraction of the inclusive dijet 
events in which the azimuthal separation of the two jets with the highest 
transverse momenta is less than a specified value for the 
parameter $\Dphimax$.
The quantity $\Rdphi$ is measured in $\ppbar$ collisions at 
$\sqrt{s}=1.96\,$TeV, as a function of the dijet rapidity interval, 
the total scalar transverse momentum, and $\Dphimax$.
The measurement uses an event sample corresponding to an integrated 
luminosity of $0.7\,$fb$^{-1}$ collected with the D0 detector at the 
Fermilab Tevatron Collider.
The results are compared to predictions of a perturbative QCD calculation 
at next-to-leading order in the strong coupling with corrections for 
non-perturbative effects.
The theory predictions describe the data, except in the
kinematic region of large dijet rapidity intervals
and large $\Dphimax$.
\end{abstract}

\maketitle


In high-energy collisions of hadrons, the production rates of particle jets
with large transverse momentum with respect to the beam direction,
$p_T$, are predicted by perturbative Quantum Chromodynamics (pQCD).
At second order in the strong coupling constant, $\as$, pQCD predicts 
only the production of dijet final states.
In the absence of higher-order radiative effects, the jet directions
are correlated in the azimuthal plane and their relative azimuthal angle 
$\Dphi = | \phi_{\rm jet1} - \phi_{\rm jet2} |$ is equal to $\pi$.
Deviations from $\pi$ (hereafter referred to as ``azimuthal decorrelations'') 
are caused by radiative processes in which additional jets are produced.
The amount of the decorrelation is directly related to the multiplicity 
and the $p_T$ carried by the additional jets.
The transition from soft to hard higher-order pQCD processes can be 
studied by examining the corresponding range of azimuthal decorrelations 
from small to large values.
This makes measurements of dijet azimuthal decorrelations an ideal 
testing ground for pQCD predictions of multijet production processes.
In pQCD, dijet azimuthal decorrelations are predicted to depend not only
on the transverse momentum of the jets, but also on the rapidity 
$\ystar = |y_{\rm jet1}-y_{\rm jet2}|/2$, obtained from the 
rapidity difference of the two leading $p_T$ jets in an event~\cite{rapidity}.
In a previous analysis of dijet azimuthal decorrelations, we measured the 
dijet differential cross section as a function of $\Dphi$, integrated over 
a fixed jet rapidity range and normalized by the inclusive dijet cross section,
for different requirements on the leading jet $p_T$~\cite{Abazov:2004hm}.
The same methodology was later used in analyses of $pp$ collision data 
at $\sqrt{s}=7\,$TeV from the CERN Large Hadron 
Collider~\cite{Khachatryan:2011zj,daCosta:2011ni}.
In all cases, dijet azimuthal decorrelations have been observed to decrease 
with increasing $p_T$; however, the combined rapidity and $p_T$ dependence 
has not yet been measured.

In this Letter, we perform a measurement of the rapidity and the $p_T$ 
dependence of dijet azimuthal decorrelations.
The analysis is based on a new quantity, $\Rdphi$, which was recently 
proposed in Ref.~\cite{Wobisch:2012au} as
\begin{equation}
 \Rdphi(H_T, y^*, \Dphimax) \, = \,
 \frac{
 \frac{d^2\sigma_{\rm dijet}(\Dphi < \Dphimax)}{dH_T \, dy^*} }%
 {\frac{d^2\sigma_{\rm dijet}(\mbox{\footnotesize inclusive}) }{dH_T \, dy^*}} \, .
 \label{eq:rdphi}
\end{equation}
The quantity $\Rdphi$ is defined as the fraction of the inclusive dijet 
cross section with a decorrelation of $\Dphi < \Dphimax$, where 
$\Dphimax$ is a parameter and $\sigma_{\rm dijet}(\mbox{inclusive})$ is 
the inclusive dijet cross section without a $\Dphi$ requirement.
It is measured as a function of $\Dphimax$, $\ystar$, and of the total 
transverse momentum $H_T$ in the event, computed as the scalar $p_T$ sum 
from all jets $i$ with $p_{Ti} > \ptmin$ and $|y_i - \yboost| < \ystarmax$ 
where $\yboost = (y_{\rm jet1} + y_{\rm jet2})/2$, $\ptmin = 30\,$GeV, 
and $\ystarmax = 2$, 
where jet1 and jet2 are the jets with the largest $p_T$ in the event.
For $\Dphimax \approx \pi$, $\Rdphi$ is sensitive to soft QCD radiation, 
while it becomes sensitive to hard higher-order QCD processes for smaller 
values of $\Dphimax$.
The phase space region $\Dphimax < 2\pi/3$ is dominated by final states 
with four or more jets.
Since $\Rdphi$ is defined as a ratio of cross sections, several experimental 
and theoretical uncertainties cancel.
In pQCD, $\Rdphi$ is computed as a ratio of three-jet and dijet cross sections 
which is (at leading order, LO) proportional to $\as$.
While dependencies on parton distribution functions (PDFs) largely cancel,  
$\Rdphi$ is sensitive to the pQCD matrix elements and to $\as$.

The measurement is performed in $\ppbar$ collisions at $\sqrt{s}= 1.96\,$TeV,
for an inclusive dijet event sample defined by the Run~II midpoint cone jet 
algorithm~\cite{run2cone} with a cone of radius $\Rcone =0.7$ in $y$ 
and $\phi$.
The dijet phase space is defined by the requirements
$p_{T1} > H_T/3$, $p_{T2} > \ptmin$, $\ystar < \ystarmax$, 
and $|\yboost| < 0.5$.
Following the proposal in Ref.~\cite{Wobisch:2012au}, 
$\Rdphi$ is measured over the 
$H_T$ range of $180$--$900$\,GeV, in three rapidity regions of $0<\ystar<0.5$, 
$0.5<\ystar<1$, and $1<\ystar<2$; and for $\Dphimax = 7\pi/8$, $5\pi/6$, 
and $3\pi/4$.
The ranges in $\ystar$ and $\yboost$, and the value of $\ptmin$ ensure that 
all jets are always within $|y|<2.5$ at $p_T$ values where the jet energy 
calibration and jet $p_T$ resolutions are known with high precision.
The requirement $p_{T1} > H_T/3$ provides a lower boundary for the leading 
jet $p_T$ in each $H_T$ bin, which (together with  $|y|<2.5$) ensures that 
the jet triggers are efficient.
The data are corrected for experimental effects and are presented at the 
``particle level,'' which includes all stable particles as defined 
in Ref.~\cite{Buttar:2008jx}.

\begin{figure*}
\includegraphics[scale=0.97]{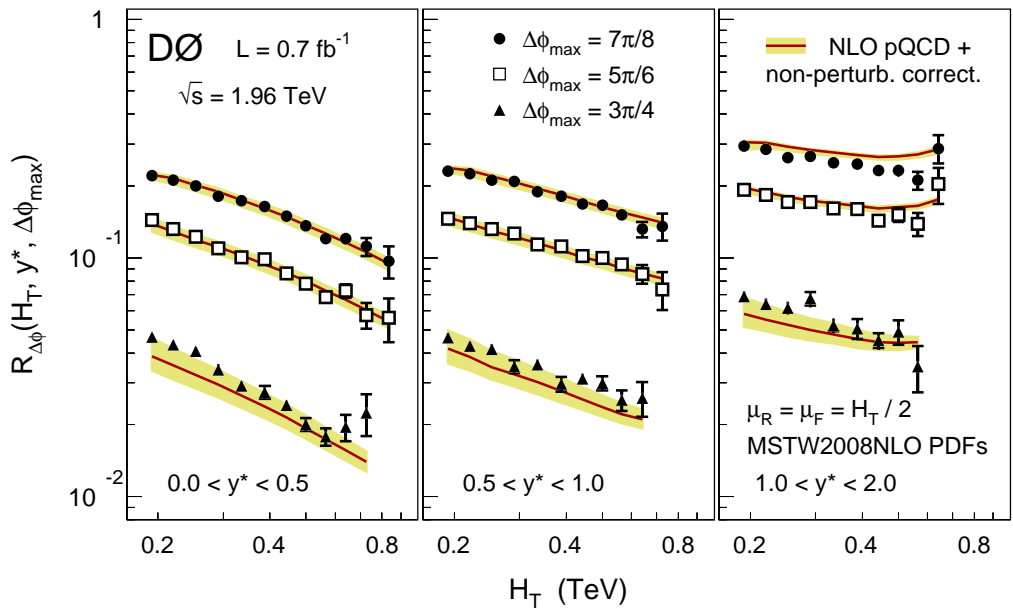}
\caption{(Color online.)
The results for $\Rdphi$ as a function of $H_T$
in three different regions of $y^*$
and for three different $\Dphimax$ requirements.
The error bars indicate the statistical and systematic 
uncertainties summed in quadrature.
The theoretical predictions are shown with their uncertainties.
\label{fig:fig1} }
\end{figure*}


A detailed description of the D0 detector is provided in Ref.~\cite{d0det}.
The event triggering and selection, jet reconstruction, and jet energy and 
momentum correction are identical to those used in recent D0 multijet 
measurements~\cite{:2009mh,Abazov:2010fr,Abazov:2011ub,:2012xib,:2012sy}.
Jets are reconstructed in the finely segmented liquid-argon sampling calorimeters 
that cover most of the solid angle.
The central calorimeter covers polar angles in the range $37$--$143^\circ$
and the two endcap calorimeters extend this coverage to within $1.7^\circ$
of the nominal beamline~\cite{d0det}.
The transition regions between the central and the endcap calorimeters
contain scintillator-based detectors to improve the energy sampling.
The jet transverse momenta are calculated using only calorimeter information
and the location of the $\ppbar$ collision.
The position of the $p\bar{p}$ interaction is determined from the tracks 
reconstructed based on data from the silicon detector and scintillating 
fiber tracker located inside a $2\,\text{T}$ solenoidal magnet~\cite{d0det}.
The position is required to be within $50$\,cm of the detector center in 
the coordinate along the beam axis, with at least three tracks pointing 
to it.
These requirements discard (7--9)\% of the events, depending on the 
trigger used.
For this measurement, events are triggered by inclusive jet triggers.
Trigger efficiencies are studied as a function of $H_T$ by comparing 
the inclusive dijet cross section in data sets obtained by triggers with 
different $p_T$ thresholds in regions where the trigger with lower 
threshold is fully efficient. 
The trigger with lowest $p_T$ threshold is shown to be fully efficient by 
studying an event sample obtained independently with a muon trigger.
In each inclusive jet $H_T$ bin, events are used from a single trigger
which has an efficiency higher than 98\%.
Requirements on the characteristics of the calorimeter clusters shower 
shapes are used to suppress the background due to electrons, 
photons, and detector noise that would otherwise mimic jets. 
The efficiency for the shower shape requirements is above $97.5\%$.
Contributions from cosmic ray events are suppressed by requiring the 
missing transverse momentum in an event to be less than 70\% (50\%) of 
the leading jet $p_T$  (before the jet energy calibration is applied)
if the latter is below (above) 100\,GeV.
The efficiency of this requirement for signal is found to be 
$>99.5\%$~\cite{:2008hua,Abazov:2011vi}.
After all selection requirements, the fraction of background events 
is below $0.1$\% for all $H_T$, as determined from distributions in 
signal and in background-enriched event samples.

The jet four-momenta reconstructed from calorimeter energy depositions
are then corrected, on average, for the response of the calorimeter, the 
net energy flow through the jet cone, additional energy from previous 
beam crossings, and multiple $p\bar{p}$ interactions in the same event, 
but not for the presence of muons and neutrinos~\cite{:2008hua,Abazov:2011vi}. 
These corrections adjust the reconstructed jet energy to the energy of 
the stable particles that enter the calorimeter except for muons and 
neutrinos.
The absolute energy calibration is determined from $Z \rightarrow \epem$ 
events and the $p_T$ imbalance in $\gamma$ + jet events in the region 
$|y| < 0.4$.
The extension to larger rapidities is derived from dijet events using 
a similar data-driven method.
In addition, corrections in the range (2--4)\% are applied that take
into account the difference in calorimeter response due to the difference 
in the fractional contributions of quark and gluon-initiated jets in the 
dijet and the $\gamma$ + jet event samples.
These corrections are determined using jets simulated with the \pythia\ 
event generator~\cite{pythia} that have been passed through a 
{\sc geant}-based detector simulation~\cite{geant}.
The total corrections of the jet four-momenta vary between 50\% and 
20\% for jet $p_T$ between 50 and 400\,GeV.
An additional correction is applied for systematic shifts in rapidity
due to detector effects~\cite{:2008hua,Abazov:2011vi}.


The procedure that corrects the distributions 
$\Rdphi(H_T, \ystar, \Dphimax)$ for experimental effects uses 
particle-level events, generated with \sherpa\ 1.1.3~\cite{sherpa} 
with MSTW2008LO PDFs~\cite{Martin:2009iq} and with 
\pythia\ 6.419~\cite{pythia} with CTEQ6.6 PDFs~\cite{Nadolsky:2008zw}
and tune QW~\cite{Albrow:2006rt}.
The jets from these events are processed by a simulation of the 
detector response which is based on parametrizations of jet $p_T$ 
resolutions and jet reconstruction efficiencies determined from data 
and of resolutions of the polar and azimuthal angles of jets, obtained 
from a detailed simulation of the detector using {\sc geant}.

\begin{figure*}
\includegraphics[scale=0.97]{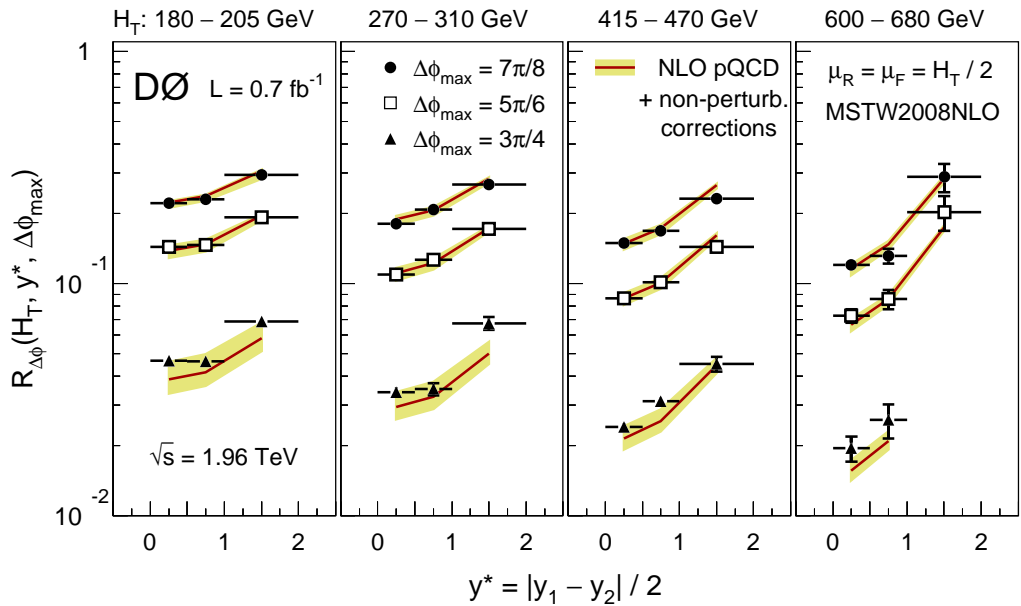}
\caption{(Color online.)
The results for $\Rdphi$ as a function of  $y^*$
in four different regions of $H_T$
and for three different $\Dphimax$ requirements.
The error bars indicate the statistical and systematic 
uncertainties summed in quadrature.
The theoretical predictions are shown with their uncertainties.
\label{fig:fig2} }
\end{figure*}

The $p_T$ resolution for jets is about 15\% at 40 GeV, decreasing 
to less than 10\% at 400 GeV.
To use the simulation to correct for experimental effects, the simulation 
must describe all relevant distributions, including the $p_T$ and 
$|y|$ distributions of the three leading $p_T$ jets, and the $\Dphi$ 
distribution.
To achieve this,
the generated events, which are used in the correction procedure,  
are weighted, based on the properties of the 
generated jets, to match these distributions in data.
The bin sizes in the $H_T$ distributions are chosen to be approximately 
twice the $H_T$ resolution.
The bin purity, defined as the fraction of all reconstructed events 
that were generated in the same bin, is above 50\% for all bins,
and only weakly dependent on $H_T$.
We then use the simulation to determine correction factors for 
experimental effects for all bins.
The correction factors are computed bin-by-bin as the ratio of $\Rdphi$ 
without and with simulation of the detector response.
These also include corrections for the energies of unreconstructed muons 
and neutrinos inside the jets.
The total correction factors for $\Rdphi$ using the weighted \pythia\ 
and \sherpa\ simulations agree typically within 1\% for $\Dphimax = 7\pi/8$ 
and $5\pi/6$ and between 1--4\%  for $\Dphimax = 3\pi/4$.
The total correction factors, defined as the average values from \pythia\ 
and \sherpa, are 0.98--1.0 for $\Dphimax = 7\pi/8$, 0.95--0.99 for 
$\Dphimax = 5\pi/6$, and 0.81--0.91 for $\Dphimax = 3\pi/4$,
with little $y^*$ dependence.
The difference between the average and the individual corrections is taken 
into account as the uncertainty attributed to the model dependence.

In total, 69 independent sources of experimental systematic uncertainties
are identified, mostly related to jet energy calibration and jet $p_T$ 
resolution.
The effects of each source are taken as fully correlated between 
all data points.
The dominant uncertainties for the $\Rdphi$ distributions are due to the 
jet energy calibration (2--5)\%, and the model dependence of the 
correction factors (1--4)\%.
Smaller contributions come from the jet $\phi$ resolution (0.5--2)\%,
from the uncertainties in systematic shifts in $y$ ($<2$\%), and the 
jet $p_T$ resolution ($<1$\%).
All other sources are negligible.
The systematic uncertainties are $2$--$3$\% for $\Dphimax = 7\pi/8$ and 
$5\pi/6$ and $3$--$5$\% for $\Dphimax = 3\pi/4$.
A detailed documentation of the results, including the individual 
contributions to the uncertainties, is provided in the 
supplementary material~\cite{suppl}.


The results for  $\Rdphi(H_T, \ystar, \Dphimax)$ are listed in 
Tables~\ref{tab:t1}--\ref{tab:t3} and displayed in Fig.~\ref{fig:fig1}
as a function of $H_T$, in different regions of $\ystar$ and for 
different $\Dphimax$.
A subset of the data points from selected $H_T$ regions are also 
shown in Fig.~\ref{fig:fig2},
where $\Rdphi$ is displayed as a function of $\ystar$ for different 
choices of $\Dphimax$.
The values of $H_T$ and $\ystar$ at which the data points are 
presented correspond to the arithmetic centers of the bins.
Figure~\ref{fig:fig1} shows that for all choices of $\Dphimax$ and 
in all $y^*$ regions, $\Rdphi$ decreases with $H_T$.
In all $y^*$ regions, the $H_T$ dependence increases towards lower 
$\Dphimax$, and for all $\Dphimax$ requirements the $H_T$ dependence 
becomes stronger for smaller $y^*$.
This implies that the $y^*$ dependence of $\Rdphi$ increases with 
increasing $H_T$, as shown in Fig.~\ref{fig:fig2}.


The theoretical predictions for $\Rdphi$ 
are obtained from a pQCD calculation in next-to-leading order (NLO),
in $\as$, with corrections for non-perturbative effects.
The latter include contributions from hadronization and the underlying event.
The non-perturbative corrections are determined using \pythia\ 6.426
with tunes  AMBT1~\cite{Diehl:2010zz} and DW~\cite{Albrow:2006rt}
which use different parton shower and underlying event models.
The hadronization correction is obtained from the ratio of $\Rdphi$
on the parton level after the parton shower 
and the particle level including all stable particles, 
both without the underlying event.
The underlying-event correction is computed from the ratio of $\Rdphi$
computed at the particle level with and without underlying event.
The total correction is given by the product of the two
individual correction factors for hadronization and the underlying event.
The total corrections vary between  $+1\%$ and $-1\%$ for tune AMBT1
and between $+1\%$ and $-3\%$ for tune DW.
The results obtained with the two tunes agree typically within
1\% and always within 3\%~\cite{Wobisch:2012au}.
The central results are taken to be the 
average values, and the uncertainty is taken to be 
half of the difference.
As a cross-check, the non-perturbative corrections are also derived with 
\herwig\/~6.520~\cite{Corcella:2000bw,Corcella:2002jc}, using default 
settings.
The \herwig\ and \pythia\ results agree typically within 0.5\%,
and always within 1\% (3\%) for $\Dphimax =7\pi/8$ and $5\pi/6$ 
(for $\Dphimax =3\pi/4$)~\cite{Wobisch:2012au}.

\begin{figure*}
\includegraphics[scale=0.94]{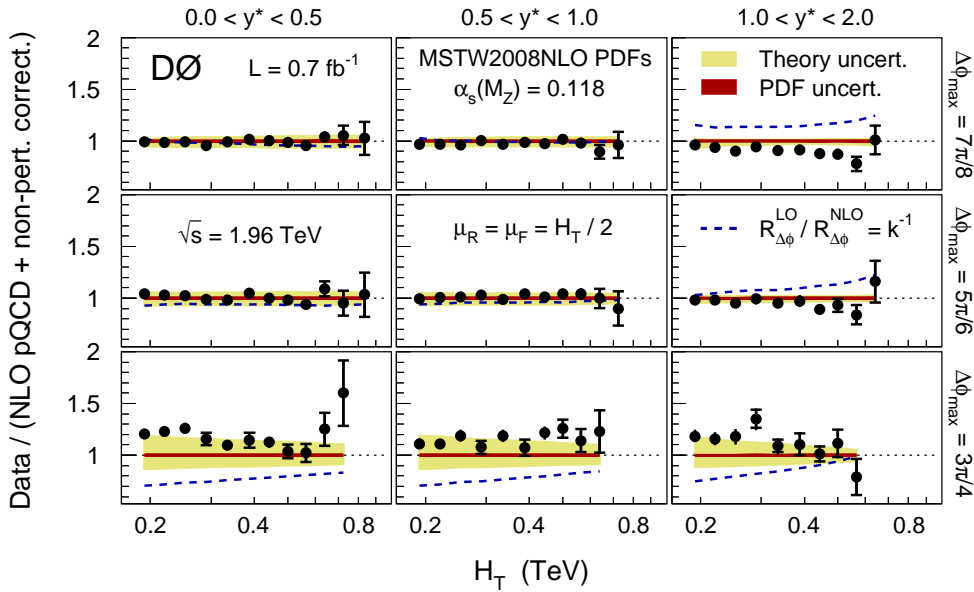}
\caption{(Color online.) 
Ratios of the results of $\Rdphi$ and the theoretical predictions
obtained for MSTW2008NLO PDFs and $\asmz=0.118$.
The ratios are shown as a function of $H_T$
in different regions of $y^*$
and for different $\Dphimax$.
The inner error bars indicate the statistical uncertainties,
and the outer error bars 
the statistical and systematic uncertainties summed in quadrature.
The theoretical uncertainty is the PDF and scale uncertainty summed
in quadrature.
Also shown is the ratio of the LO and NLO pQCD predictions
which is the inverse of the NLO $k$-factor.
\label{fig:fig3}}
\end{figure*}

The NLO (LO) pQCD prediction for $\Rdphi$ is computed as the ratio
of the NLO (LO) predictions for the numerator and the denominator.
The NLO prediction for the numerator (denominator) is obtained from 
an $\ord(\as^4)$ ($\ord(\as^3)$) cross section calculation.
These results are computed using 
{\sc fastnlo}~\cite{Kluge:2006xs,Wobisch:2011ij} based on 
{\sc nlojet++}~\cite{Nagy:2003tz,Nagy:2001fj}, in the $\overline{\mbox{MS}}$ 
scheme~\cite{Bardeen:1978yd} for five active quark flavors.
The calculations use the next-to-leading logarithmic (two-loop) 
approximation of the renormalization group equation and 
$\as(M_Z)=0.118$ in the matrix elements and the PDFs, which is close to 
the current world average value of $0.1184\pm0.0007$~\cite{PDG2012}.
The MSTW2008NLO PDFs~\cite{Martin:2009iq} are used, and the central 
choice $\mu_0$ for the renormalization and factorization scales is 
$\mur = \muf = \mu_0 = H_T/2$, which is identical to $\mu_0 = p_T$ 
for inclusive jet and dijet production at LO.
The theoretical predictions are overlaid on the data in 
Figs.~\ref{fig:fig1} and~\ref{fig:fig2}, and some properties are 
displayed in Fig.~\ref{fig:fig3}.
The PDF uncertainties are computed using the up and down variations 
of the 20 orthogonal PDF uncertainty eigenvectors, corresponding to 
the 68\% C.L., as provided by MSTW2008NLO.
The PDF uncertainties are typically 1\%, and never larger than 2\%.
The $\Rdphi$ results obtained with the CT10~\cite{Lai:2010vv} and 
NNPDFv2.1~\cite{Ball:2011mu} PDF parametrizations agree with those 
for MSTW2008NLO within 2\%.
The theoretical uncertainties are dominated by the uncertainties 
of the pQCD calculations due to the $\mur$ and $\muf$ dependencies.
These are computed as the relative changes of the results due to 
independent variations of both scales between $\mu_0 / 2 $ and 
$2\mu_0$, with the restriction of $0.5 \le \mur / \muf \le 2.0$.
The uncertainties from the scale dependence are $4$--$6$\% for 
$\Dphimax = 7\pi/8$ and $5\pi/6$, and $6$--$20$\% for 
$\Dphimax = 3\pi/4$, decreasing with $H_T$.
In addition to the scale dependence, the NLO $k$-factors provide 
additional information on the convergence of the perturbative expansion, 
and therefore on the possible size of missing higher order contributions.
The NLO $k$-factors are computed as the ratio of the NLO and the LO 
predictions for $\Rdphi$ ($k = \Rdphi^{\rm NLO} / \Rdphi^{\rm LO}$).
Figure~\ref{fig:fig3} shows the inverse of the NLO $k$-factors
and their dependence on $y^*$ and $\Dphimax$.


Ratios of data and the theoretical predictions are displayed 
in Fig.~\ref{fig:fig3} as a function of $H_T$ in all regions 
of $y^*$ and $\Dphimax$.
To quantify the agreement, $\chi^2$ values are determined that 
compare data and theory, taking into account the correlations between 
all uncertainties.
The $\chi^2$ definition is the same that was used in our recent 
$\as$ determinations~\cite{:2012xib,Abazov:2009nc}.
Table~\ref{tab:consistency} displays the $\chi^2$ values for all $H_T$ bins 
within each of the nine kinematic regions in $y^*$ and $\Dphimax$.
The results are shown for three different choices of $\mur$ and $\muf$,
including the central choice $\mur=\muf=H_T/2$ and the combined lower 
and upper variations, $H_T/4$ and $H_T$.
The following discussion distinguishes between the three different kinematic 
regions, which are given by $\Dphimax=3\pi/4$, by $y^* > 1$, and by
$y^*<1$ with $\Dphimax=7\pi/8$ or $5\pi/6$.

The region of large azimuthal decorrelations, $\Dphimax = 3\pi/4$, 
is challenging for the theoretical predictions, since it receives 
large contributions from four-jet final states.
These are only modeled at LO by the $\ord(\as^4)$ calculation for the 
numerator of $\Rdphi$, which causes the large NLO $k$-factors (up to 1.5)
and the large scale dependence (up to 21\%), seen in Fig.~\ref{fig:fig3}.
In this kinematic region, the central theoretical predictions are 
consistently below the data (often by 15--25\%).
Within the large scale uncertainty, however, they agree with the 
data, as the $\chi^2$ values for the lower scale choice $H_T/4$
are all consistent with the expectations based on the number of 
degrees of freedom ($\Ndof$, which corresponds here to the number 
of data points), of $\chi^2 = \Ndof \pm \sqrt{2 \,\Ndof}$.

In the kinematic region $y^* > 1$, the theoretical predictions exhibit 
a different $H_T$ dependence as compared to lower $y^*$, as seen in 
Fig.~\ref{fig:fig1}.
While at lower $y^*$ the predicted $H_T$ dependence of the $\Rdphi$ 
distributions is monotonically decreasing, the $H_T$ distributions 
for $y^* > 1$ have a local minimum around $\approx 0.5\,$TeV above 
which $\Rdphi$ increases.
For $\Dphimax = 5\pi/6$, the theoretical predictions give an adequate 
description of the data.
For $\Dphimax = 7\pi/8$, however, the predicted $H_T$ dependence differs
from that of the measured $\Rdphi$ distribution, as quantified by the 
large $\chi^2$ regardless of the scale choice.
This is the only kinematic region in ($\Dphimax$, $y^*$) for which 
the NLO $k$-factor is consistently below unity (0.89--0.81) over the 
entire $H_T$ range.
This may indicate a poor convergence of the perturbative expansion.

The perturbative expansion works best in the kinematic regions of 
$0<y^*<0.5$ and $0.5<y^*<1.0$, where the scale dependence is small ($<6\%$)
and the NLO $k$-factors are above unity but small ($1.00 < k <1.06$).
In all of those regions, the theoretical predictions give a good 
description of the data.

\begin{table}[t]
\centering
\caption{\label{tab:t1}
The results for $\Rdphi$ with their relative uncertainties 
for $\Dphimax=7\pi/8$.
}
\begin{ruledtabular}
\begin{tabular}{r r r c r r}
 \multicolumn{1}{c}{$H_T$}  &  \multicolumn{1}{c}{$y^*$} & \multicolumn{1}{c}{$\Rdphi$} & \multicolumn{1}{c}{Stat.~uncert.} & \multicolumn{2}{c}{Syst.~uncert.} \\
 \multicolumn{1}{c}{(GeV)} &   & & \multicolumn{1}{c}{(percent)} & \multicolumn{2}{c}{(percent)} \\
\hline
 $180$--$205$  &  $ 0.0$--$ 0.5$  &  $ 2.216 \times 10^{-1}$  &  \phantom{1}$\pm  0.9$  &   $+  2.8$  &  $-  3.0$ \\
 $205$--$235$  &  $ 0.0$--$ 0.5$  &  $ 2.116 \times 10^{-1}$  &  \phantom{1}$\pm  1.1$  &   $+  2.6$  &  $-  2.6$ \\
 $235$--$270$  &  $ 0.0$--$ 0.5$  &  $ 2.000 \times 10^{-1}$  &  \phantom{1}$\pm  1.5$  &   $+  2.5$  &  $-  2.3$ \\
 $270$--$310$  &  $ 0.0$--$ 0.5$  &  $ 1.811 \times 10^{-1}$  &  \phantom{1}$\pm  2.2$  &   $+  2.2$  &  $-  2.2$ \\
 $310$--$360$  &  $ 0.0$--$ 0.5$  &  $ 1.731 \times 10^{-1}$  &  \phantom{1}$\pm  1.5$  &   $+  2.0$  &  $-  2.1$ \\
 $360$--$415$  &  $ 0.0$--$ 0.5$  &  $ 1.641 \times 10^{-1}$  &  \phantom{1}$\pm  2.4$  &   $+  1.9$  &  $-  1.9$ \\
 $415$--$470$  &  $ 0.0$--$ 0.5$  &  $ 1.491 \times 10^{-1}$  &  \phantom{1}$\pm  1.5$  &   $+  1.9$  &  $-  1.9$ \\
 $470$--$530$  &  $ 0.0$--$ 0.5$  &  $ 1.359 \times 10^{-1}$  &  \phantom{1}$\pm  2.4$  &   $+  1.9$  &  $-  1.9$ \\
 $530$--$600$  &  $ 0.0$--$ 0.5$  &  $ 1.206 \times 10^{-1}$  &  \phantom{1}$\pm  3.2$  &   $+  2.0$  &  $-  2.0$ \\
 $600$--$680$  &  $ 0.0$--$ 0.5$  &  $ 1.204 \times 10^{-1}$  &  \phantom{1}$\pm  5.0$  &   $+  2.2$  &  $-  2.1$ \\
 $680$--$770$  &  $ 0.0$--$ 0.5$  &  $ 1.114 \times 10^{-1}$  &  \phantom{1}$\pm  8.8$  &   $+  2.4$  &  $-  2.3$ \\
 $770$--$900$  &  $ 0.0$--$ 0.5$  &  $ 9.699 \times 10^{-2}$  &  $\pm 15.4$  &   $+  2.5$  &  $-  2.3$ \\
\hline
 $180$--$205$  &  $ 0.5$--$ 1.0$  &  $ 2.311 \times 10^{-1}$  &  \phantom{1}$\pm  0.9$  &   $+  2.9$  &  $-  3.3$ \\
 $205$--$235$  &  $ 0.5$--$ 1.0$  &  $ 2.252 \times 10^{-1}$  &  \phantom{1}$\pm  1.2$  &   $+  2.8$  &  $-  2.9$ \\
 $235$--$270$  &  $ 0.5$--$ 1.0$  &  $ 2.115 \times 10^{-1}$  &  \phantom{1}$\pm  1.6$  &   $+  2.6$  &  $-  2.5$ \\
 $270$--$310$  &  $ 0.5$--$ 1.0$  &  $ 2.085 \times 10^{-1}$  &  \phantom{1}$\pm  2.3$  &   $+  2.4$  &  $-  2.3$ \\
 $310$--$360$  &  $ 0.5$--$ 1.0$  &  $ 1.888 \times 10^{-1}$  &  \phantom{1}$\pm  1.7$  &   $+  2.2$  &  $-  2.2$ \\
 $360$--$415$  &  $ 0.5$--$ 1.0$  &  $ 1.808 \times 10^{-1}$  &  \phantom{1}$\pm  2.8$  &   $+  2.1$  &  $-  2.1$ \\
 $415$--$470$  &  $ 0.5$--$ 1.0$  &  $ 1.686 \times 10^{-1}$  &  \phantom{1}$\pm  1.8$  &   $+  2.1$  &  $-  2.0$ \\
 $470$--$530$  &  $ 0.5$--$ 1.0$  &  $ 1.662 \times 10^{-1}$  &  \phantom{1}$\pm  2.8$  &   $+  2.1$  &  $-  2.1$ \\
 $530$--$600$  &  $ 0.5$--$ 1.0$  &  $ 1.517 \times 10^{-1}$  &  \phantom{1}$\pm  3.9$  &   $+  2.2$  &  $-  2.1$ \\
 $600$--$680$  &  $ 0.5$--$ 1.0$  &  $ 1.318 \times 10^{-1}$  &  \phantom{1}$\pm  7.3$  &   $+  2.4$  &  $-  2.1$ \\
 $680$--$770$  &  $ 0.5$--$ 1.0$  &  $ 1.356 \times 10^{-1}$  &  $\pm 13.0$  &   $+  2.5$  &  $-  2.2$ \\
\hline
 $180$--$205$  &  $ 1.0$--$ 2.0$  &  $ 2.934 \times 10^{-1}$  &  \phantom{1}$\pm  1.0$  &   $+  3.8$  &  $-  4.4$ \\
 $205$--$235$  &  $ 1.0$--$ 2.0$  &  $ 2.850 \times 10^{-1}$  &  \phantom{1}$\pm  1.3$  &   $+  3.7$  &  $-  4.0$ \\
 $235$--$270$  &  $ 1.0$--$ 2.0$  &  $ 2.634 \times 10^{-1}$  &  \phantom{1}$\pm  2.0$  &   $+  3.6$  &  $-  3.4$ \\
 $270$--$310$  &  $ 1.0$--$ 2.0$  &  $ 2.667 \times 10^{-1}$  &  \phantom{1}$\pm  2.9$  &   $+  3.3$  &  $-  2.9$ \\
 $310$--$360$  &  $ 1.0$--$ 2.0$  &  $ 2.502 \times 10^{-1}$  &  \phantom{1}$\pm  2.3$  &   $+  3.1$  &  $-  2.6$ \\
 $360$--$415$  &  $ 1.0$--$ 2.0$  &  $ 2.468 \times 10^{-1}$  &  \phantom{1}$\pm  4.0$  &   $+  3.1$  &  $-  2.6$ \\
 $415$--$470$  &  $ 1.0$--$ 2.0$  &  $ 2.317 \times 10^{-1}$  &  \phantom{1}$\pm  3.0$  &   $+  3.1$  &  $-  2.7$ \\
 $470$--$530$  &  $ 1.0$--$ 2.0$  &  $ 2.320 \times 10^{-1}$  &  \phantom{1}$\pm  5.2$  &   $+  3.0$  &  $-  2.7$ \\
 $530$--$600$  &  $ 1.0$--$ 2.0$  &  $ 2.116 \times 10^{-1}$  &  \phantom{1}$\pm  8.6$  &   $+  2.8$  &  $-  2.8$ \\
 $600$--$680$  &  $ 1.0$--$ 2.0$  &  $ 2.875 \times 10^{-1}$  &  $\pm 13.7$  &   $+  2.7$  &  $-  2.8$ \\
\end{tabular}
\end{ruledtabular}
\end{table}

\begin{table}[t]
\centering
\caption{\label{tab:t2}
The results for $\Rdphi$ with their relative uncertainties 
for $\Dphimax=5\pi/6$.
}
\begin{ruledtabular}
\begin{tabular}{r r r c r r}
 \multicolumn{1}{c}{$H_T$}  &  \multicolumn{1}{c}{$y^*$} & \multicolumn{1}{c}{$\Rdphi$} & \multicolumn{1}{c}{Stat.~uncert.} & \multicolumn{2}{c}{Syst.~uncert.} \\
 \multicolumn{1}{c}{(GeV)} &   & & \multicolumn{1}{c}{(percent)} & \multicolumn{2}{c}{(percent)} \\
 \hline
 $180$--$205$  &  $ 0.0$--$ 0.5$  &  $ 1.439 \times 10^{-1}$  &  \phantom{1}$\pm  1.1$  &  $+  2.8$  &  $-  2.6$ \\
 $205$--$235$  &  $ 0.0$--$ 0.5$  &  $ 1.325 \times 10^{-1}$  &  \phantom{1}$\pm  1.4$  &  $+  2.5$  &  $-  2.5$ \\
 $235$--$270$  &  $ 0.0$--$ 0.5$  &  $ 1.223 \times 10^{-1}$  &  \phantom{1}$\pm  2.0$  &  $+  2.3$  &  $-  2.3$ \\
 $270$--$310$  &  $ 0.0$--$ 0.5$  &  $ 1.097 \times 10^{-1}$  &  \phantom{1}$\pm  3.0$  &  $+  2.1$  &  $-  2.1$ \\
 $310$--$360$  &  $ 0.0$--$ 0.5$  &  $ 1.007 \times 10^{-1}$  &  \phantom{1}$\pm  2.1$  &  $+  2.0$  &  $-  2.0$ \\
 $360$--$415$  &  $ 0.0$--$ 0.5$  &  $ 9.851 \times 10^{-2}$  &  \phantom{1}$\pm  3.3$  &  $+  2.0$  &  $-  1.9$ \\
 $415$--$470$  &  $ 0.0$--$ 0.5$  &  $ 8.635 \times 10^{-2}$  &  \phantom{1}$\pm  2.1$  &  $+  2.0$  &  $-  2.0$ \\
 $470$--$530$  &  $ 0.0$--$ 0.5$  &  $ 7.821 \times 10^{-2}$  &  \phantom{1}$\pm  3.2$  &  $+  2.0$  &  $-  2.0$ \\
 $530$--$600$  &  $ 0.0$--$ 0.5$  &  $ 6.832 \times 10^{-2}$  &  \phantom{1}$\pm  4.3$  &  $+  2.1$  &  $-  2.1$ \\
 $600$--$680$  &  $ 0.0$--$ 0.5$  &  $ 7.262 \times 10^{-2}$  &  \phantom{1}$\pm  6.6$  &  $+  2.2$  &  $-  2.3$ \\
 $680$--$770$  &  $ 0.0$--$ 0.5$  &  $ 5.760 \times 10^{-2}$  &  $\pm 12.5$  &  $+  2.3$  &  $-  2.4$ \\
 $770$--$900$  &  $ 0.0$--$ 0.5$  &  $ 5.600 \times 10^{-2}$  &  $\pm 20.7$  &  $+  2.6$  &  $-  2.7$ \\
\hline
 $180$--$205$  &  $ 0.5$--$ 1.0$  &  $ 1.463 \times 10^{-1}$  &  \phantom{1}$\pm  1.4$  &  $+  3.2$  &  $-  2.9$ \\
 $205$--$235$  &  $ 0.5$--$ 1.0$  &  $ 1.396 \times 10^{-1}$  &  \phantom{1}$\pm  1.6$  &  $+  2.6$  &  $-  2.6$ \\
 $235$--$270$  &  $ 0.5$--$ 1.0$  &  $ 1.317 \times 10^{-1}$  &  \phantom{1}$\pm  2.2$  &  $+  2.4$  &  $-  2.5$ \\
 $270$--$310$  &  $ 0.5$--$ 1.0$  &  $ 1.263 \times 10^{-1}$  &  \phantom{1}$\pm  3.1$  &  $+  2.2$  &  $-  2.3$ \\
 $310$--$360$  &  $ 0.5$--$ 1.0$  &  $ 1.139 \times 10^{-1}$  &  \phantom{1}$\pm  2.3$  &  $+  2.2$  &  $-  2.2$ \\
 $360$--$415$  &  $ 0.5$--$ 1.0$  &  $ 1.117 \times 10^{-1}$  &  \phantom{1}$\pm  3.6$  &  $+  2.1$  &  $-  2.1$ \\
 $415$--$470$  &  $ 0.5$--$ 1.0$  &  $ 1.016 \times 10^{-1}$  &  \phantom{1}$\pm  2.4$  &  $+  2.1$  &  $-  2.1$ \\
 $470$--$530$  &  $ 0.5$--$ 1.0$  &  $ 9.993 \times 10^{-2}$  &  \phantom{1}$\pm  3.8$  &  $+  2.1$  &  $-  2.1$ \\
 $530$--$600$  &  $ 0.5$--$ 1.0$  &  $ 9.414 \times 10^{-2}$  &  \phantom{1}$\pm  5.1$  &  $+  2.2$  &  $-  2.2$ \\
 $600$--$680$  &  $ 0.5$--$ 1.0$  &  $ 8.566 \times 10^{-2}$  &  \phantom{1}$\pm  9.2$  &  $+  2.2$  &  $-  2.3$ \\
 $680$--$770$  &  $ 0.5$--$ 1.0$  &  $ 7.369 \times 10^{-2}$  &  $\pm 18.2$  &  $+  2.3$  &  $-  2.5$ \\
\hline
 $180$--$205$  &  $ 1.0$--$ 2.0$  &  $ 1.926 \times 10^{-1}$  &  \phantom{1}$\pm  1.3$  &  $+  4.0$  &  $-  3.5$ \\
 $205$--$235$  &  $ 1.0$--$ 2.0$  &  $ 1.840 \times 10^{-1}$  &  \phantom{1}$\pm  1.8$  &  $+  3.7$  &  $-  3.5$ \\
 $235$--$270$  &  $ 1.0$--$ 2.0$  &  $ 1.709 \times 10^{-1}$  &  \phantom{1}$\pm  2.6$  &  $+  3.3$  &  $-  3.4$ \\
 $270$--$310$  &  $ 1.0$--$ 2.0$  &  $ 1.716 \times 10^{-1}$  &  \phantom{1}$\pm  3.9$  &  $+  3.0$  &  $-  3.2$ \\
 $310$--$360$  &  $ 1.0$--$ 2.0$  &  $ 1.611 \times 10^{-1}$  &  \phantom{1}$\pm  3.0$  &  $+  3.0$  &  $-  3.1$ \\
 $360$--$415$  &  $ 1.0$--$ 2.0$  &  $ 1.600 \times 10^{-1}$  &  \phantom{1}$\pm  5.2$  &  $+  3.0$  &  $-  3.1$ \\
 $415$--$470$  &  $ 1.0$--$ 2.0$  &  $ 1.436 \times 10^{-1}$  &  \phantom{1}$\pm  4.0$  &  $+  3.0$  &  $-  3.0$ \\
 $470$--$530$  &  $ 1.0$--$ 2.0$  &  $ 1.518 \times 10^{-1}$  &  \phantom{1}$\pm  6.7$  &  $+  2.9$  &  $-  3.1$ \\
 $530$--$600$  &  $ 1.0$--$ 2.0$  &  $ 1.391 \times 10^{-1}$  &  $\pm 11.0$  &  $+  2.8$  &  $-  3.1$ \\
 $600$--$680$  &  $ 1.0$--$ 2.0$  &  $ 2.034 \times 10^{-1}$  &  $\pm 17.2$  &  $+  2.9$  &  $-  3.2$ \\
\end{tabular}
\end{ruledtabular}
\end{table}

\begin{table}[t]
\centering
\caption{\label{tab:t3}
The results for $\Rdphi$ with their relative uncertainties 
for $\Dphimax=3\pi/4$.
}
\begin{ruledtabular}
\begin{tabular}{r r r c r r}
 \multicolumn{1}{c}{$H_T$}  &  \multicolumn{1}{c}{$y^*$} & \multicolumn{1}{c}{$\Rdphi$} & \multicolumn{1}{c}{Stat.~uncert.} & \multicolumn{2}{c}{Syst.~uncert.} \\
 \multicolumn{1}{c}{(GeV)} &   & & \multicolumn{1}{c}{(percent)} & \multicolumn{2}{c}{(percent)} \\
\hline
 $180$--$205$  &  $ 0.0$--$ 0.5$  &  $ 4.659 \times 10^{-2}$  &  \phantom{1}$\pm  2.0$  &  $+  2.8$  &  $-  2.7$ \\
 $205$--$235$  &  $ 0.0$--$ 0.5$  &  $ 4.339 \times 10^{-2}$  &  \phantom{1}$\pm  2.6$  &  $+  2.7$  &  $-  2.5$ \\
 $235$--$270$  &  $ 0.0$--$ 0.5$  &  $ 4.055 \times 10^{-2}$  &  \phantom{1}$\pm  3.5$  &  $+  2.3$  &  $-  2.2$ \\
 $270$--$310$  &  $ 0.0$--$ 0.5$  &  $ 3.405 \times 10^{-2}$  &  \phantom{1}$\pm  5.4$  &  $+  2.2$  &  $-  2.2$ \\
 $310$--$360$  &  $ 0.0$--$ 0.5$  &  $ 2.913 \times 10^{-2}$  &  \phantom{1}$\pm  4.0$  &  $+  2.2$  &  $-  2.2$ \\
 $360$--$415$  &  $ 0.0$--$ 0.5$  &  $ 2.733 \times 10^{-2}$  &  \phantom{1}$\pm  6.2$  &  $+  2.2$  &  $-  2.3$ \\
 $415$--$470$  &  $ 0.0$--$ 0.5$  &  $ 2.419 \times 10^{-2}$  &  \phantom{1}$\pm  4.0$  &  $+  2.2$  &  $-  2.4$ \\
 $470$--$530$  &  $ 0.0$--$ 0.5$  &  $ 2.008 \times 10^{-2}$  &  \phantom{1}$\pm  6.3$  &  $+  2.1$  &  $-  2.6$ \\
 $530$--$600$  &  $ 0.0$--$ 0.5$  &  $ 1.780 \times 10^{-2}$  &  \phantom{1}$\pm  8.4$  &  $+  2.3$  &  $-  2.8$ \\
 $600$--$680$  &  $ 0.0$--$ 0.5$  &  $ 1.953 \times 10^{-2}$  &  $\pm 12.6$  &  $+  2.7$  &  $-  3.1$ \\
 $680$--$770$  &  $ 0.0$--$ 0.5$  &  $ 2.241 \times 10^{-2}$  &  $\pm 19.8$  &  $+  3.8$  &  $-  3.5$ \\
\hline
 $180$--$205$  &  $ 0.5$--$ 1.0$  &  $ 4.620 \times 10^{-2}$  &  \phantom{1}$\pm  2.6$  &  $+  3.9$  &  $-  3.9$ \\
 $205$--$235$  &  $ 0.5$--$ 1.0$  &  $ 4.261 \times 10^{-2}$  &  \phantom{1}$\pm  3.0$  &  $+  3.2$  &  $-  3.1$ \\
 $235$--$270$  &  $ 0.5$--$ 1.0$  &  $ 4.152 \times 10^{-2}$  &  \phantom{1}$\pm  3.9$  &  $+  2.7$  &  $-  2.7$ \\
 $270$--$310$  &  $ 0.5$--$ 1.0$  &  $ 3.510 \times 10^{-2}$  &  \phantom{1}$\pm  6.0$  &  $+  2.6$  &  $-  2.8$ \\
 $310$--$360$  &  $ 0.5$--$ 1.0$  &  $ 3.578 \times 10^{-2}$  &  \phantom{1}$\pm  4.1$  &  $+  2.6$  &  $-  2.9$ \\
 $360$--$415$  &  $ 0.5$--$ 1.0$  &  $ 2.962 \times 10^{-2}$  &  \phantom{1}$\pm  7.1$  &  $+  2.7$  &  $-  2.9$ \\
 $415$--$470$  &  $ 0.5$--$ 1.0$  &  $ 3.107 \times 10^{-2}$  &  \phantom{1}$\pm  4.3$  &  $+  2.7$  &  $-  2.9$ \\
 $470$--$530$  &  $ 0.5$--$ 1.0$  &  $ 2.984 \times 10^{-2}$  &  \phantom{1}$\pm  6.9$  &  $+  2.6$  &  $-  2.9$ \\
 $530$--$600$  &  $ 0.5$--$ 1.0$  &  $ 2.532 \times 10^{-2}$  &  \phantom{1}$\pm  9.8$  &  $+  2.6$  &  $-  3.2$ \\
 $600$--$680$  &  $ 0.5$--$ 1.0$  &  $ 2.587 \times 10^{-2}$  &  $\pm 16.7$  &  $+  3.1$  &  $-  3.4$ \\
\hline
 $180$--$205$  &  $ 1.0$--$ 2.0$  &  $ 6.873 \times 10^{-2}$  &  \phantom{1}$\pm  2.5$  &  $+  4.8$  &  $-  3.9$ \\
 $205$--$235$  &  $ 1.0$--$ 2.0$  &  $ 6.402 \times 10^{-2}$  &  \phantom{1}$\pm  3.3$  &  $+  4.5$  &  $-  4.1$ \\
 $235$--$270$  &  $ 1.0$--$ 2.0$  &  $ 6.169 \times 10^{-2}$  &  \phantom{1}$\pm  4.6$  &  $+  4.3$  &  $-  4.5$ \\
 $270$--$310$  &  $ 1.0$--$ 2.0$  &  $ 6.741 \times 10^{-2}$  &  \phantom{1}$\pm  6.4$  &  $+  4.1$  &  $-  4.8$ \\
 $310$--$360$  &  $ 1.0$--$ 2.0$  &  $ 5.218 \times 10^{-2}$  &  \phantom{1}$\pm  5.5$  &  $+  4.2$  &  $-  4.7$ \\
 $360$--$415$  &  $ 1.0$--$ 2.0$  &  $ 5.049 \times 10^{-2}$  &  \phantom{1}$\pm  9.5$  &  $+  4.3$  &  $-  4.5$ \\
 $415$--$470$  &  $ 1.0$--$ 2.0$  &  $ 4.505 \times 10^{-2}$  &  \phantom{1}$\pm  7.2$  &  $+  4.3$  &  $-  4.4$ \\
 $470$--$530$  &  $ 1.0$--$ 2.0$  &  $ 4.899 \times 10^{-2}$  &  $\pm 11.9$  &  $+  4.1$  &  $-  4.7$ \\
 $530$--$600$  &  $ 1.0$--$ 2.0$  &  $ 3.504 \times 10^{-2}$  &  $\pm 22.0$  &  $+  3.6$  &  $-  5.6$ \\
\end{tabular}
\end{ruledtabular}
\end{table}


In summary, the first measurement of the combined rapidity and $p_T$ 
dependence of dijet azimuthal decorrelations is presented.
The measurement is based on the recently proposed quantity $\Rdphi$,
which probes dijet azimuthal decorrelations in a novel way.
It is measured in $\ppbar$ collisions at $\sqrt{s}=1.96\,$TeV
as a function of the total transverse 
momentum $H_T$, the rapidity $\ystar$, and the parameter $\Dphimax$.
For all values of $\Dphimax$ and at fixed $H_T$, dijet azimuthal 
decorrelations increase with $\ystar$, while they decrease with $H_T$ 
over most of the $H_T$ range at fixed $\ystar$.
Predictions of NLO pQCD, corrected for non-perturbative effects,
give a good description of the data, except in the kinematic region 
of large dijet rapidity intervals $\ystar > 1$ and small decorrelations 
$\Dphimax = 7\pi/8$.

\begin{table}[b]
\centering
\caption{\label{tab:consistency}
The $\chi^2$ values between data and theory for MSTW2008PDFs and 
$\asmz = 0.118$ and for different choices of $\mur$ and $\muf$. 
The results are shown for each of the nine kinematic regions, 
defined by the $y^*$ and $\Dphimax$ requirements, 
combining all $H_T$ bins inside those regions.
}
\begin{ruledtabular}
\begin{tabular}{cccrrr}
$y^*$ & $\Dphimax$ & $\Ndof$ & \multicolumn{3}{c}{$\chi^2$ for $\mur = \muf =$} \\
range &  &  & $H_T/4$ & $H_T/2$ & $H_T$ \\
\hline
0.0--0.5 & 7$\pi$/8 & 12  &  15.1 &  7.1 &  12.7  \\
0.0--0.5 & 5$\pi$/6 & 12  &  15.7 &  10.9 &  20.9  \\
0.0--0.5 & 3$\pi$/4 & 11  &  13.1 &  44.2 &  104.5  \\
\hline
0.5--1.0 & 7$\pi$/8 & 11  &  11.8 &  6.9 &  8.6  \\
0.5--1.0 & 5$\pi$/6 & 11  &  5.6 &  4.0 &  12.6   \\
0.5--1.0 & 3$\pi$/4 & 10  &  15.4 &  26.9 &  60.2  \\
\hline
1.0--2.0 & 7$\pi$/8 & 10  &  29.7 &  24.4 &  19.8  \\
1.0--2.0 & 5$\pi$/6 & 10  &  9.3 &  10.8 &  10.7  \\
1.0--2.0 & 3$\pi$/4 & \phantom{1}9 &  10.3 &  23.1 &  45.5  \\
\end{tabular}
\end{ruledtabular}
\end{table}

\input acknowledgement.tex
\end{document}

%% file: author_list.tex
\affiliation{LAFEX, Centro Brasileiro de Pesquisas F\'{i}sicas, Rio de Janeiro, Brazil}
\affiliation{Universidade do Estado do Rio de Janeiro, Rio de Janeiro, Brazil}
\affiliation{Universidade Federal do ABC, Santo Andr\'e, Brazil}
\affiliation{University of Science and Technology of China, Hefei, People's Republic of China}
\affiliation{Universidad de los Andes, Bogot\'a, Colombia}
\affiliation{Charles University, Faculty of Mathematics and Physics, Center for Particle Physics, Prague, Czech Republic}
\affiliation{Czech Technical University in Prague, Prague, Czech Republic}
\affiliation{Center for Particle Physics, Institute of Physics, Academy of Sciences of the Czech Republic, Prague, Czech Republic}
\affiliation{Universidad San Francisco de Quito, Quito, Ecuador}
\affiliation{LPC, Universit\'e Blaise Pascal, CNRS/IN2P3, Clermont, France}
\affiliation{LPSC, Universit\'e Joseph Fourier Grenoble 1, CNRS/IN2P3, Institut National Polytechnique de Grenoble, Grenoble, France}
\affiliation{CPPM, Aix-Marseille Universit\'e, CNRS/IN2P3, Marseille, France}
\affiliation{LAL, Universit\'e Paris-Sud, CNRS/IN2P3, Orsay, France}
\affiliation{LPNHE, Universit\'es Paris VI and VII, CNRS/IN2P3, Paris, France}
\affiliation{CEA, Irfu, SPP, Saclay, France}
\affiliation{IPHC, Universit\'e de Strasbourg, CNRS/IN2P3, Strasbourg, France}
\affiliation{IPNL, Universit\'e Lyon 1, CNRS/IN2P3, Villeurbanne, France and Universit\'e de Lyon, Lyon, France}
\affiliation{III. Physikalisches Institut A, RWTH Aachen University, Aachen, Germany}
\affiliation{Physikalisches Institut, Universit\"at Freiburg, Freiburg, Germany}
\affiliation{II. Physikalisches Institut, Georg-August-Universit\"at G\"ottingen, G\"ottingen, Germany}
\affiliation{Institut f\"ur Physik, Universit\"at Mainz, Mainz, Germany}
\affiliation{Ludwig-Maximilians-Universit\"at M\"unchen, M\"unchen, Germany}
\affiliation{Fachbereich Physik, Bergische Universit\"at Wuppertal, Wuppertal, Germany}
\affiliation{Panjab University, Chandigarh, India}
\affiliation{Delhi University, Delhi, India}
\affiliation{Tata Institute of Fundamental Research, Mumbai, India}
\affiliation{University College Dublin, Dublin, Ireland}
\affiliation{Korea Detector Laboratory, Korea University, Seoul, Korea}
\affiliation{CINVESTAV, Mexico City, Mexico}
\affiliation{Nikhef, Science Park, Amsterdam, the Netherlands}
\affiliation{Radboud University Nijmegen, Nijmegen, the Netherlands}
\affiliation{Joint Institute for Nuclear Research, Dubna, Russia}
\affiliation{Institute for Theoretical and Experimental Physics, Moscow, Russia}
\affiliation{Moscow State University, Moscow, Russia}
\affiliation{Institute for High Energy Physics, Protvino, Russia}
\affiliation{Petersburg Nuclear Physics Institute, St. Petersburg, Russia}
\affiliation{Instituci\'{o} Catalana de Recerca i Estudis Avan\c{c}ats (ICREA) and Institut de F\'{i}sica d'Altes Energies (IFAE), Barcelona, Spain}
\affiliation{Uppsala University, Uppsala, Sweden}
\affiliation{Lancaster University, Lancaster LA1 4YB, United Kingdom}
\affiliation{Imperial College London, London SW7 2AZ, United Kingdom}
\affiliation{The University of Manchester, Manchester M13 9PL, United Kingdom}
\affiliation{University of Arizona, Tucson, Arizona 85721, USA}
\affiliation{University of California Riverside, Riverside, California 92521, USA}
\affiliation{Florida State University, Tallahassee, Florida 32306, USA}
\affiliation{Fermi National Accelerator Laboratory, Batavia, Illinois 60510, USA}
\affiliation{University of Illinois at Chicago, Chicago, Illinois 60607, USA}
\affiliation{Northern Illinois University, DeKalb, Illinois 60115, USA}
\affiliation{Northwestern University, Evanston, Illinois 60208, USA}
\affiliation{Indiana University, Bloomington, Indiana 47405, USA}
\affiliation{Purdue University Calumet, Hammond, Indiana 46323, USA}
\affiliation{University of Notre Dame, Notre Dame, Indiana 46556, USA}
\affiliation{Iowa State University, Ames, Iowa 50011, USA}
\affiliation{University of Kansas, Lawrence, Kansas 66045, USA}
\affiliation{Kansas State University, Manhattan, Kansas 66506, USA}
\affiliation{Louisiana Tech University, Ruston, Louisiana 71272, USA}
\affiliation{Northeastern University, Boston, Massachusetts 02115, USA}
\affiliation{University of Michigan, Ann Arbor, Michigan 48109, USA}
\affiliation{Michigan State University, East Lansing, Michigan 48824, USA}
\affiliation{University of Mississippi, University, Mississippi 38677, USA}
\affiliation{University of Nebraska, Lincoln, Nebraska 68588, USA}
\affiliation{Rutgers University, Piscataway, New Jersey 08855, USA}
\affiliation{Princeton University, Princeton, New Jersey 08544, USA}
\affiliation{State University of New York, Buffalo, New York 14260, USA}
\affiliation{University of Rochester, Rochester, New York 14627, USA}
\affiliation{State University of New York, Stony Brook, New York 11794, USA}
\affiliation{Brookhaven National Laboratory, Upton, New York 11973, USA}
\affiliation{Langston University, Langston, Oklahoma 73050, USA}
\affiliation{University of Oklahoma, Norman, Oklahoma 73019, USA}
\affiliation{Oklahoma State University, Stillwater, Oklahoma 74078, USA}
\affiliation{Brown University, Providence, Rhode Island 02912, USA}
\affiliation{University of Texas, Arlington, Texas 76019, USA}
\affiliation{Southern Methodist University, Dallas, Texas 75275, USA}
\affiliation{Rice University, Houston, Texas 77005, USA}
\affiliation{University of Virginia, Charlottesville, Virginia 22904, USA}
\affiliation{University of Washington, Seattle, Washington 98195, USA}
\author{V.M.~Abazov} \affiliation{Joint Institute for Nuclear Research, Dubna, Russia}
\author{B.~Abbott} \affiliation{University of Oklahoma, Norman, Oklahoma 73019, USA}
\author{B.S.~Acharya} \affiliation{Tata Institute of Fundamental Research, Mumbai, India}
\author{M.~Adams} \affiliation{University of Illinois at Chicago, Chicago, Illinois 60607, USA}
\author{T.~Adams} \affiliation{Florida State University, Tallahassee, Florida 32306, USA}
\author{G.D.~Alexeev} \affiliation{Joint Institute for Nuclear Research, Dubna, Russia}
\author{G.~Alkhazov} \affiliation{Petersburg Nuclear Physics Institute, St. Petersburg, Russia}
\author{A.~Alton$^{a}$} \affiliation{University of Michigan, Ann Arbor, Michigan 48109, USA}
\author{A.~Askew} \affiliation{Florida State University, Tallahassee, Florida 32306, USA}
\author{S.~Atkins} \affiliation{Louisiana Tech University, Ruston, Louisiana 71272, USA}
% ...
%
%
\collaboration{The D0 Collaboration\footnote{with visitors from
$^{a}$Augustana College, Sioux Falls, SD, USA,
$^{b}$The University of Liverpool, Liverpool, UK,
$^{c}$UPIITA-IPN, Mexico City, Mexico,
$^{d}$DESY, Hamburg, Germany,
$^{e}$SLAC, Menlo Park, CA, USA,
$^{f}$University College London, London, UK,
$^{g}$Centro de Investigacion en Computacion - IPN, Mexico City, Mexico,
$^{h}$ECFM, Universidad Autonoma de Sinaloa, Culiac\'an, Mexico
and
$^{i}$Universidade Estadual Paulista, S\~ao Paulo, Brazil.
}} \noaffiliation
\vskip 0.25cm

%% file: acknowledgement.tex
%
We thank the staffs at Fermilab and collaborating institutions,
and acknowledge support from the
DOE and NSF (USA);
CEA and CNRS/IN2P3 (France);
MON, NRC KI and RFBR (Russia);
CNPq, FAPERJ, FAPESP and FUNDUNESP (Brazil);
DAE and DST (India);
Colciencias (Colombia);
CONACyT (Mexico);
NRF (Korea);
FOM (The Netherlands);
STFC and the Royal Society (United Kingdom);
MSMT and GACR (Czech Republic);
BMBF and DFG (Germany);
SFI (Ireland);
The Swedish Research Council (Sweden);
and
CAS and CNSF (China).